\documentstyle[prb,aps,epsf,preprint]{revtex}
\draft

\begin{document}
\title{Magic Structures of Helical Multi-shell Zirconium Nanowires}

\author{Baolin Wang $^{1,2}$, Guanghou Wang $^1$ \thanks{Corresponding author, E-mail: ghwang@nju.edu.cn, Fax: +86-25-3595535.}, Jijun Zhao $^{3}$}
\address{$^1$ National Laboratory of Solid State Microstructures and Department of Physics, Nanjing University, Nanjing 210093, China \\
$^2$ Department of Physics, Huaiyin Teachers Colloge, Jiangsu 223001, P.R.China \\
$^3$ Department of Physics and Astronomy, University of North Carolina at Chapel Hill, Chapel Hill, NC 27599\\}

\maketitle

\begin{abstract} 

The structures of free-standing zirconium nanowires with 0.6$-$2.8 nm in diameter are systematically studied by using genetic algorithm simulations with a tight-binding many body potential. Several multi-shell growth sequences with cylindrical structures are obtained. These multi-shell structures are composed of coaxial atomic shells with the three- and four-strands helical, centered pentagonal and hexagonal, and parallel double-chain-core curved surface epitaxy. Under the same growth sequence, the numbers of atomic strands in inner- and outer-shell show even-odd coupling and usually differ by five. The size and structure dependence of angular correlation functions and vibrational properties of zirconium nanowire are also discussed.

\end{abstract}
\pacs{61.46.+w, 68.65.+g, 73.61.-r}

In the past decade, there has been tremendous interests on ultrathin metal nanowires from both fundamental low-dimensional physics and technological applications such as nanoelectronic devices \cite {1,2,3,4,5,6,7,8,9,10,11,12,13,14}. Most of the previous studies are based on the tip-surface contact \cite{1,2,3,4,5} or mechanically controllable break junctions \cite{6,7}, which can be seen as very short metal nanowire. In recent experiments, Takayanagi's group has successfully fabricated stable gold wires with various diameter of sufficient length suspended between two stands \cite{8,9,10}, and the helical multi-shell structures are observed in those ultrathin wires \cite{10}. On the theoretical side, the noncrystalline structures, melting behavior, and electronic properties of ultrathin free standing Pb, Al, Ag nanowires have been investigated by Tosatti and co-workers \cite{11,12,13,14}. By using molecular dynamics-based (MD) genetic algorithm (GA) simulations, our group has studied structural, vibrational, electronic, magnetic properties of gold, titanium, and rhodium nanowires \cite{15,16,17}. However, the current knowledge on the detailed structural characters and growth sequences of the metal nanowires is still quite limited. In this paper, we report systematical multi-shell structural growth sequence of zirconium nanowires and their vibrational properties. 

In our simulations, the zirconium nanowires with diameters from 0.6 to 2.8 nm are modeled by a supercell with one-dimensional (1D) periodical boundary condition along the wire axis direction.  For the most zirconium nanowires studied, the length of 1-D supercell is chosen to be 1.413 nm, which is a reasonable compromise between discovering the helicity in 1D direction and avoiding the nanowires breaking into clusters upon relaxation. To match the periodicity of the multi-shell structures with pentagonal symmetry, several different supercell lengths, i.e., 1.177 nm for S5-1, S5-2 wires (see Fig.1), 1.648 nm for S5-3, S5-4 wires, have been used. The ``effective diameter'', which can be controlled by adjusting the number of atoms in the supercell, is used to denote the cross section area of the nanowires \cite{12,15,16,17}. To search the most stable structure of nanowire, we adopt the genetic algorithm (GA) based on MD relaxation \cite{15,16,17}. The vibrational densities of states can be then calculated by diagonalizing the dynamical matrix for the optimized configurations.

The interaction between zirconium atoms is described by a tight binding (TB) many body potential \cite{18}, which has successfully reproduced the high-temperature hcp-bcc transition in bulk zirconium \cite{18,19}. To further check the validity of TB potential in low-dimensional systems with reduced coordination number (CN), we have performed density functional calculations on small zirconium clusters (Zr$_7$, Zr$_{13}$, Zr$_{15}$, Zr$_{19}$) and bulk zirconium solid by using a DMol package \cite{20} within local density approximation (LDA). In both TB and LDA calculations, the cluster structures are optimized and the similar equilibrium configurations are obtained. Typically, the discrepancy for interatomic distance obtained from tight-binding potential and LDA calculation are within 0.1 {\AA}. The comparison of LDA and TB results on the binding energies of small clusters with equilibrium structures are given in Table I. The satisfactory agreement between LDA and TB results on Zr clusters in Table I demonstrates the validity of the TB potential in the other low-dimensional systems like nanowire. 

\begin{table}
Table I. Comparison of binding energies from tight-binding (TB) potential and density functional calculation with local density approximation (LDA) for the small Zr$_n$ clusters ($n=7, 13, 15, 19$). The binding energies of clusters are given by the percentage of bulk cohesive energy and the average coordination number (CN) are presented. The TB results on larger Zr$_{38}$ and Zr$_{55}$ clusters are also included in the Table. The equilibrium configurations for the Zr$_7$, Zr$_{13}$, Zr$_{15}$, Zr$_{19}$, Zr$_{38}$, Zr$_{55}$ are pentagonal bipyramid, icosahedron, hexagonal anti-prism, double icosahedron, truncated octahedron, icosahedron respectively. 
\begin{center}
\begin{tabular}{ccccccc}
        & Zr$_7$ & Zr$_{13}$ &Zr$_{15}$ &  Zr$_{19}$ &Zr$_{38}$ & Zr$_{55}$ \\ \hline
CN      & 4.57   & 6.46      &  6.67    &   7.16     &  7.58    & 8.51      \\
LDA     &65.8$\%$& 72.8$\%$  & 76.7$\%$ & 77.2$\%$   &          &           \\
TB      &67.5$\%$& 75.6$\%$  & 76.3$\%$ & 78.3$\%$   & 82.8$\%$ & 85.3$\%$  \\
\end{tabular}
\end{center}
\end{table}

\begin{figure}
\centerline{
\epsfxsize=5.0in \epsfbox{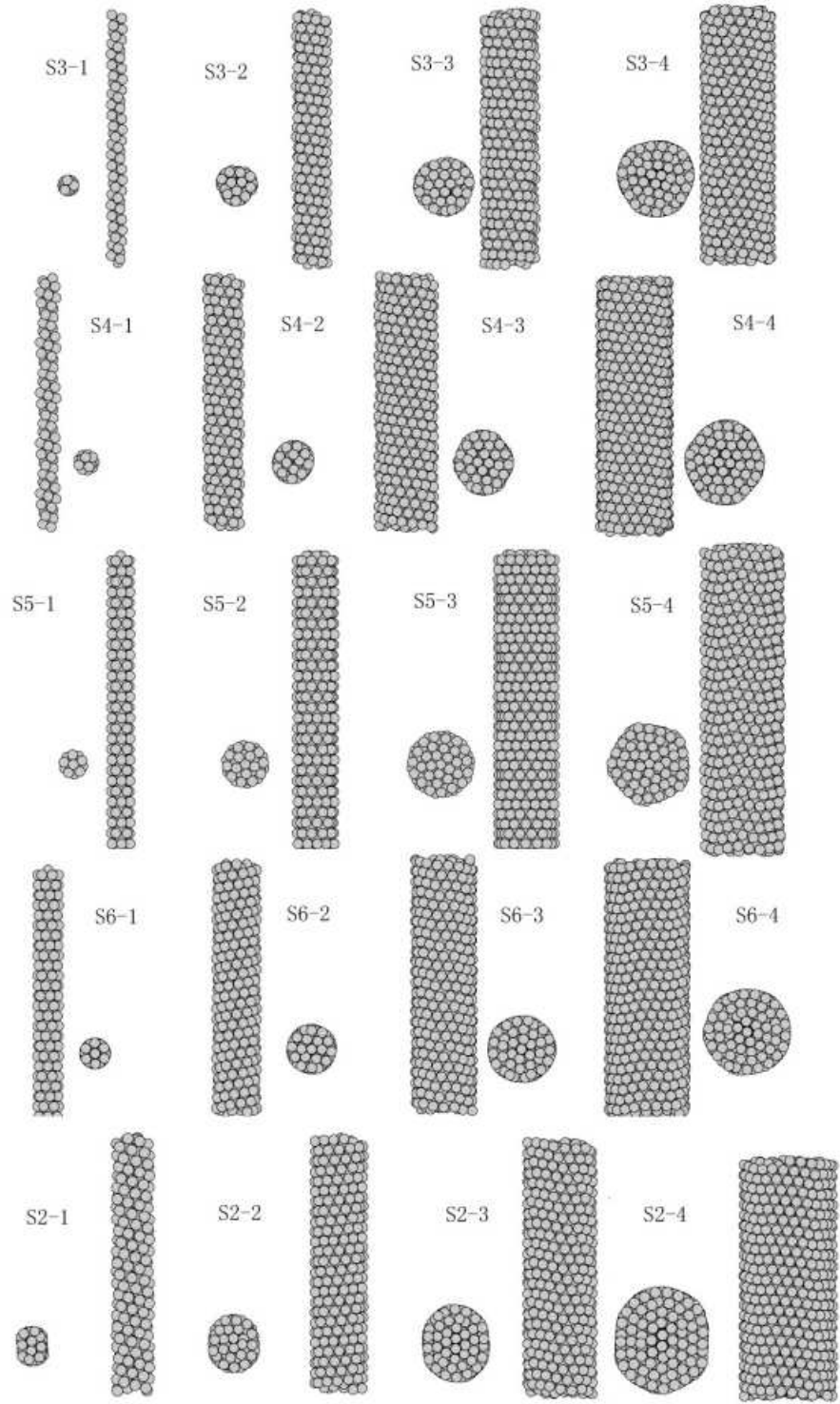}
}
\caption{Morphologies of structural growth sequences in zirconium nanowires, the cross sections in right part for S4-x,S6-x and left one for S3-x, S5-x, and S2-x (x=1-4).}
\end{figure}

From the GA simulation, we have obtained numerous isomers for the zirconium nanowires. Among these structural candidates, we found that the ones with closed-shell structures are usually the lowest-energy configurations at a given wire diameter. The energy difference between the lowest-energy closed-shell structures and other open-shell structures is about 0.005eV/atom. For example, the binding energy of helical closed-shell S6-4 wire (see Fig.1) is -5.829 eV/atom, while a low-energy non-helical open-shell configuration at the same size has a binding energy -5.824 eV/atom. In our simulations, we have tested different supercell length and found that the multi-shell arrangements and the helicity of magic closed-shell nanowires do not sensitively depend on the chosen supercell length. On the other hand, those non-magic open-shell wires are either crystalline-like or irregular non-crystalline and have higher binding energies. These results indicate the relative stability of magic closed-shell structure and the possibility of observing these metal nanowires experimentally.

Shown in Fig.1 are five kinds of growth sequences of zirconium nanowires obtained from GA optimization. These multi-shell nanowires are composed of coaxial cylindrical atomic shells. Each shell is usually formed by a number of atomic strands winding up helically side-by-side. The pitches of the helix for the outer and inner shells are different. The lateral surface of the cylindrical shell forms a triangular atomic network. Such helical multi-shell structures have been theoretically predicted for Al, Pb \cite{12}, Au \cite{15}, Ti\cite{16}, and Rh nanowires \cite{17}, and experimentally observed in Au nanowires \cite{10}. In Fig.1, we introduce the index S3-x, S4-x, S5-x, S6-x, and S2-x (x=1-4) to characterize the multi-shell (x: number of shells) growth patterns based on trigonal (S3-x), tetragonal (S4-x), centered pentagonal (S5-x), centered hexagonal (S6-x), and double-chain-core (S2-x) packing, respectively. These twenty structures shown in Fig.1 represent five kind of continuous structural growth sequences of zirconium nanowires. Such systematic multi-shell structural growth patterns in metal nanowires have never been reported before.

With the notation of $n$-$n1$-$n2$-$n3$-$n4$, we can describe the closed-shell magic nanowires consisting of coaxial shells, each of which is composed of $n, n1, n2, n3$, and $n4$ helical atomic strands from outer to inner ($n>n1>n2>n3>n4$) \cite{10,15,16}. As shown in Fig.1, wires S3-1(3), S3-2(8-3), S3-3(13-8-3), and S3-4(18-13-8-3) constitute the sequence of trigonal multi-shell packing. The structure of the thinnest wire S3-1 (diameter D=0.574 nm) is formed by three-strands packing helically (see Fig.1). Similarly, wires S4-1(4), S4-2(9-4), S4-3(14-9-4), and S4-4(19-14-9-4) constitute growth patterns with one-, two-, three-, and four-shell tetragonal packing. This sequence starts from the single-shell wire S4-1 with four-strands helical packing. In both the trigonal and tetragonal multi-shell growth sequence, the outer and inner shells differ by five helical atomic strands. 

In contrast to the helical structures found in S3-x, S4-x wires, the wires S5-1(5-1), S5-2(10-5-1), and S5-3(15-10-5-1) under the centered pentagonal multi-shell sequence are not helical, while S5-4(20-15-10-5-1) come back to helical. These structures can be viewed as deformed icosahedral packing. Wires S6-1(6-1), S6-2(11-6-1), S6-3(16-11-6-1), and S6-4(21-16-11-6-1) constitute the sequence of centered hexagonal packing. They are the most ideal cylindrical configurations in all the structures studied. In the double-chain-core sequence of wires S2-1(8-2), S2-2(13-8-2), S2-3(18-13-8-2), and S2-4(23-18-13-8-2), the central core consists a nearly parallel dimer-chain. The surrounding shells are made up of 8, 13, 18, and 23 identical atomic strands respectively. Each strand helically wound with a linear pitch around the cylinder. The differences between the number of atomic strands in outer and inner shells are $5,5,5, $ and $6$, respectively. The thickest wire in Fig.1 is double-chain-core S2-4 wire with D=2.787 nm. It is worthy noted that, in most growth sequence of zirconium nanowires, the number ($n, n^{\prime}$) of atomic stands that forming the inner and outer shells typically have even-odd coupling with $n-n^{\prime} =5$, with exception at S2-1 and S5-1 wires. These behaviors are similar to those obtained from previous experiment on suspended gold nanowires \cite{10}. In the experiment, the number $n, n^{\prime}$ of atomic strands in inner and outer shells also exhibit even-odd coupling with $n-n^{\prime} =7$, which can be related to the string tension. 

\newpage
\begin{table}[tbp]
Table II. Structures, diameters (D), average binding energies per atom (E), and average coordination numbers (CN) for the multi-shell zirconium nanowires with different growth pattern. The shell numbers $n$-$n1$-n2-n3-n4 for these wires is given in parenthesis. Bulk cohesive energy from present TB potential is $E_b$=-6.167 eV/atom.
\begin{center}
\begin{tabular}{ccccccc}
Structures(n-n1-n2-n3-n4) & D(nm) & $-$E(eV/atom) & CN &  &  &  \\ \hline
S3-1(3) & 0.574 & 4.588 & 6.000 &  &  &  \\ 
S3-2(8-3) & 1.168 & 5.434 & 9.069 &  &  &  \\ 
S3-3(13-8-3) & 1.715 & 5.667 & 9.824 &  &  &  \\ 
S3-4(18-13-8-3) & 2.275 & 5.786 & 10.036 &  &  &  \\ 
S4-1(4) & 0.651 & 4.782 & 7.000 &  &  &  \\ 
S4-2(9-4) & 1.197 & 5.447 & 9.097 &  &  &  \\ 
S4-3(14-9-4) & 1.802 & 5.688 & 9.797 &  &  &  \\ 
S4-4(19-14-9-4) & 2.401 & 5.807 & 10.131 &  &  &  \\ 
S5-1(5-1) & 0.823 & 5.131 & 8.667 &  &  &  \\ 
S5-2(10-5-1) & 1.344 & 5.526 & 9.813 &  &  &  \\ 
S5-3(15-10-5-1) & 1.932 & 5.735 & 10.142 &  &  &  \\ 
S5-4(20-15-10-5-1) & 2.400 & 5.792 & 10.157 &  &  &  \\ 
S6-1(6-1) & 0.908 & 5.235 & 8.857 &  &  &  \\ 
S6-2(11-6-1) & 1.426 & 5.564 & 9.136 &  &  &  \\ 
S6-3(16-11-6-1) & 2.002 & 5.746 & 9.800 &  &  &  \\ 
S6-4(21-16-11-6-1) & 2.571 & 5.829 & 10.121 &  &  &  \\ 
S2-1(8-2) & 1.044 & 5.334 & 8.222 &  &  &  \\ 
S2-2(13-8-2) & 1.609 & 5.629 & 9.673 &  &  &  \\ 
S2-3(18-13-8-2) & 2.159 & 5.761 & 9.859 &  &  &  \\ 
S2-4(23-18-13-8-2) & 2.787 & 5.850 & 10.206 &  &  & 
\end{tabular}
\end{center}
\end{table}

The structures, diameters, coordination number, and binding energies of the zirconium nanowires in equilibrium structures are summarized in Table II. The average binding energies $E$ and coordination numbers of zirconium nanowires generally increase as the diameter increase. The largest binding energy of the atoms in the thickest wires (S2-4) reach about $95\%$ of the bulk value (-6.167 eV/atom) \cite{18} and that of the thinnest one (S3-1) is about $75\%$. If we plot the average binding energies $E$ as function of $1/D$, all the nanowires from different growth patterns fit well to the same linear dependence as: $E=E_b-\lambda/D$, where coefficient $\lambda$=0.85 eV and the wire diameter D is in unit of nm. The linear dependence with 1/D can be understood by the surface effect \cite{12}. The binding energies of all these multi-shell wires fitting to the same line indicate that they are almost equally stable.

\begin{figure}
\vspace{1.0in}
\centerline{
\epsfxsize=5.5in \epsfbox{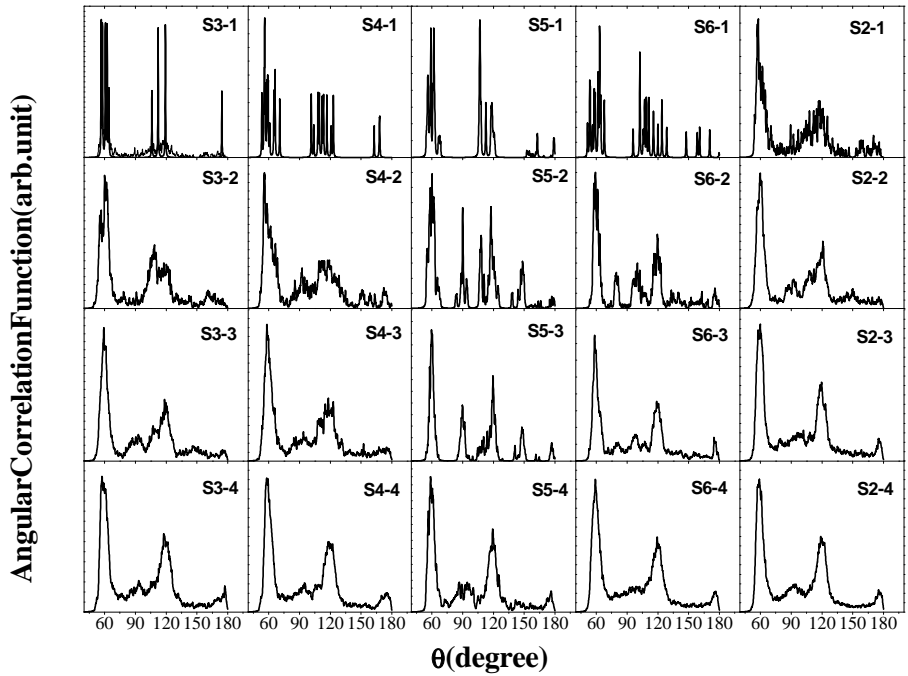}
}
\vspace{-1.5in}
\caption{Angular correlation functions of zirconium nanowires.
}
\end{figure}

To illustrate the structural characters of zirconium nanowires, the angular correlation functions (ACFs) of these nanowires are calculated and presented in Fig.2. For most structures, two major peaks in the ACFs located around $60^o$ and $120^o$ are found, which are related to the hexagonal packing. The broad distributions of ACFs in addition to the two major peaks demonstrate that noncrystalline multi-shell structures do not have any definite bond angle. From the ACFs shown in Fig.2, it is clearly seen that discrete feature of the ACF peaks gradually disappear as the wire diameter increases and the bulk character gradually come out. On the other hand, there are considerable differences between the angular correlation functions for the different growth sequences in Fig.2. For instance, the features related to deformed icosahedron can be observed in S5-x (x=1-4) sequence, while perfect icosahedral bond angles are 63.4$^o$, 116.6$^o$, and 180$^o$ \cite{12}.

\begin{figure}
\vspace{1.0in}
\centerline{
\epsfxsize=5.5in \epsfbox{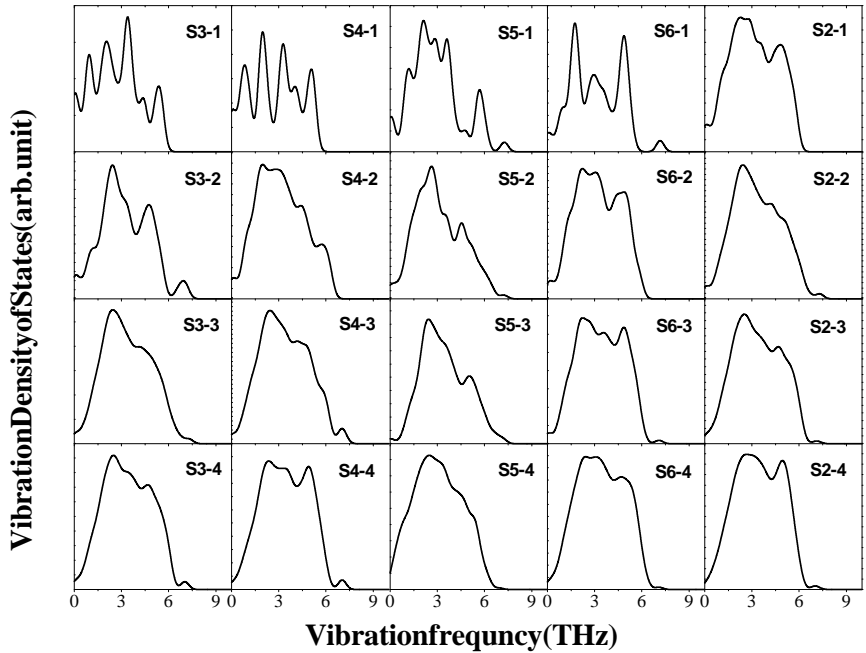}
}
\vspace{-1.5in}
\caption{Vibrational densities of states of zirconium clusters.}
\end{figure}

Based on the optimized structures of zirconium nanowires, we now their vibrational properties. The vibrational densities of states of zirconium nanowires in Fig.3 exhibit remarkable dependence on the wire diameter and the corresponding growth classes. Except wires S3-1 and S4-1, the position and shape of the major vibrational peak located around $2.1-2.7$ THz do not sensitively change with the size and growth sequence of wires. This behavior is similar to that found for gold nanowires in our previous work \cite{15}. The first major peak around $2.1-2.7$ THz can be related to the normal vibrational model of hcp bulk zirconium\cite{21}. In an earlier study \cite{21}, the frequency distribution of the bulk zirconium evaluated at room temperature range between 1.2 $-$ 6.1 THz. The first bulk frequency peak were found at 2.1$-$2.3 THz, which is comparable to the major peak ($2.1-2.7$ THz) in zirconium nanowires. Different from the first major peak, although the positions of second peak ($\sim$ 5 THz) found for the nanowires do not substantially change, its shape and height sensitively vary with the size and growth sequence. Such differences might be understood by the different vibrational couplings between the atomic shells for the nanowires with various sizes and growth classes. In another words, the vibrational modes of zirconium nanowires sensitively depend on the specific growth patterns of the constituent atomic strands and diameters of nanowires. The vibrational bands of the thinner wires (S3-1, S4-1) are rather discrete and molecule-like. Some high frequency modes around $7.5THz$ are found in S2-x, S3-x and S4-x sequences. These high frequency vibration modes can be attributed to the contraction of interatomic distance in those wires, which will be further illustrated in the following discussion on zirconium clusters.

\begin{figure}
\vspace{-0.5in}
\centerline{
\epsfxsize=3.0in \epsfbox{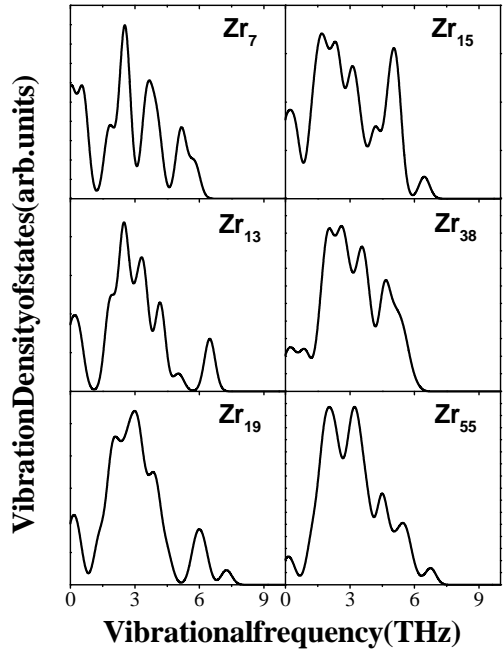}
}
\vspace{-0.4in}
\caption{Vibrational densities of states of zirconium clusters.}
\end{figure}

In order to further analyze the relations between the vibrational spectra and atomic structures, we have calculated the vibrational densities of states for the zirconium clusters (Zr$_7$, Zr$_{13}$, Zr$_{15}$ , Zr$_{19}$, Zr$_{38}$, and Zr$_{55}$) (Fig.4). These vibrational densities of states of clusters are comparable to those of the bulk and nanowires. Because of the structural similarity, the shapes of vibrational densities of states for Zr$_{13}$ and Zr$_{15}$ are very closed to those of the pentagonal S5-1 and hexagonal S6-1 wires, respectively. The vibrational models of the clusters can be also understood from their atomic structures. The high frequencies are due to the large force constants, which are the consequence of the contraction of the interatomic distance in the clusters. For instance, the distances between surface atoms and central atom in an icosahedral Zr$_{13}$ cluster (about 2.96$-$2.98 {\AA}) is smaller than these between surface atoms (3.10 $-$ 3.14 {\AA}), which leads to the high frequency about 6.5 THz. Similar effect are observed in Zr$_{15}$, Zr$_{19}$, Zr$_{38}$, and Zr$_{55}$. The appearance of some low-frequency modes for zirconium clusters may be attributed to the atoms weakly bonded to the cluster. 

From the above discussions, we can make the following conclusions. Multi-shell cylindrical structural growth sequences are found for the zirconium nanowires up to about 3 nm. The number of atomic strands in outer and inner shells exhibit even-odd and the differences between them is usually five for most growth sequence of zirconium nanowires. The angular correlation functions and vibrational properties are related to the wire diameter and growth pattern. The helical multi-shell metal nanowires represent a novel structural form of matter. They are similar but not identical to clusters, which is yet to be fully understood \cite{14,22}. In principle, the mechanical, thermal, electronic, magnetic, and transport properties of such nanowires may sensitively depend on their atomic structures and helicity. The future studies in these directions are under way. We expect future experiments on transition metal nanowires to validate our arguments.

\ \\
B. L. Wang and G. H. Wang would like to thank financial support from
National Nature Science Foundation of China (No:$29890210,10023001$).


\begin{references}
\bibitem{1}  J. I. Pascual, J. Mendez, J. Gomez-Herrero, A. M. Baro, N.
Garcia, V. T. Binh, Phys. Rev. Lett. 71, 1852(1993).

\bibitem{2}  N. Agrait, J. G.Rodrigo, and S. Vieira, Phys. Rev. B47.
12345(1993).

\bibitem{3}  J. M. Krans, C. J. Muller, I. K. Yanson, Th. C. M.Govaert, R.
Hesper, J. M. van Ruitenbeek, Phys. Rev .B48, 14721(1993).

\bibitem{4}  L. Olesen, E. Laegsgaard, I. Stensgaard, F. Besenbacher, J.
Schiotz, P. Stoltze, K. W. Jacobsen, J. K. Norskov, Phys. Rev. Lett. 72,
2251(1994).

\bibitem{5}  J. I. Pascual, J. Mendez, J. Gomez-Herrero, A. M.Baro, N.
Garcia, U. Landman, W. D. Luedtke, E. N. Bogachek, H. P. Cheng, Science 267,
1793(1995).

\bibitem{6}  J. M. Krans, J. M. van Ruitenbeek, V. V. Fisun, J. K. Yanson,
L. J. de Jongh, Nature 375, 767(1995).

\bibitem{7}  N. Agrait, G. Rubio. S. Vieira, Phys. Rev. Lett. 74,
3995(1995); G. Rubio, N. Agrait, S. Vieira, Phys. Rev. Lett. 76, 2302(1996).

\bibitem{8}  Y. Kondo, K. Takayanagi, Phys .Rev. Lett. 79, 3455(1997).

\bibitem{9}  H. Ohnishim, Y. Kondo, K. Takayanagi, Nature 395, 780(1998).

\bibitem{10}  K. Kondo, K. Takayanagi, Science 289, 606(2000).

\bibitem{11}  O. Gulseren, F. Ercolessi, E. Tosatti, Phys. Rev. B51,
7377(1995).

\bibitem{12}  O. Gulseren, F. Ercolessi, E. Tosatti, Phys. Rev. Lett. 80,
3775(1998).

\bibitem{13}  F. Di Tolla, A. Dal Corse, J. A. Torres, E.Tosatti, Surf. Sci.
456, 947(2000).

\bibitem{14}  E. Tosatti, S. Prestipino, S. Kostlmeier, A. Dal Corso, F. Di
Tolla, Science 291, 288(2001).

\bibitem{15}  B. L. Wang, S. Y. Yin, G. H. Wang, A. Buldum, J. J. Zhao,
Phys. Rev. Lett. 86, 2046(2001).

\bibitem{16}  B. L. Wang, S. Y. Yin, G. H. Wang, J. J. Zhao, J. Phys.:
Condens. Matter 13, 403(2001).

\bibitem{17}  B. L. Wang, Y. Ren, H. Q. Sun, G. H. Wang, J. J. Zhao,
unpublished.

\bibitem{18}  F. Cleri, V. Rosato, Phys. Rev. B48, 22(1993).

\bibitem{19}  F. Willaime, C. Massobrio, Phys. Rev. Lett. 63, 2244(1989).

\bibitem{20} DMOL is a density functional theory (DFT) package based atomic basis distributed by Accelrys (B.Delley, J.Chem.Phys.{\bf 92}, 508(1990)). 

\bibitem{21}  C. Stassis, J. Zarestky, D. Arch, O. D. McMasters, and B. N.
Harmon, Phys.Rev.B{\bf 18}, 2632(1978).

\bibitem{22}  E. Tosatti, S. Prestipino, Science 289, 561(2000).

\end{references}
\end{document}